\documentclass{aa}
\usepackage{epsfig}

\begin{document}
\title{Molecular gas at intermediate redshifts}
\titlerunning{Molecular gas at intermediate redshifts}
\author{Nissim Kanekar \inst{1}\thanks{nissim@ncra.tifr.res.in}, 
Jayaram N Chengalur\inst{1}\thanks{chengalu@ncra.tifr.res.in}}
\authorrunning{Kanekar \& Chengalur }
\institute{National Centre for Radio Astrophysics, Post Bag 3, Ganeshkhind, Pune 411 007}
\date{Received mmddyy/ accepted mmddyy}
\offprints{Nissim Kanekar}
\abstract
{  We present Giant Metrewave Radio Telescope (GMRT) observations of OH absorption 
   in B3~1504+377 ($z \sim 0.673$) and PKS~1413+135 ($z \sim 0.247$). OH has 
   now been detected in absorption towards four intermediate redshift
   systems, viz. the lensing galaxies towards B~0218+357 ($z \sim 0.685$; Kanekar 
   et al. 2001) and 1830-211 ($z \sim 0.886$; Chengalur et al. 1999), in addition 
   to the two systems listed above.  All four systems also give rise to well 
   studied millimetre wavelength molecular line absorption from a host of 
   molecules, including HCO$^+$. Comparing our OH data with these millimetre line 
   transitions, we find that the linear correlation between $N_{\rm OH}$ and 
   $N_{\rm HCO^+}$ found
   in molecular clouds in the Milky Way (Liszt \& Lucas 1996) persists out to 
   $z \sim 1$. It has been suggested (Liszt \& Lucas 1999) that OH is a good 
   tracer of ${\rm H_2}$, with $N_{\rm H_2}/N_{\rm OH} \approx 10^7$ under a 
   variety of physical conditions. We use this relationship to estimate 
   $N_{\rm H_2}$ in these absorbers. The estimated $N_{\rm H_2}$ is $\ga 10^{22}$ 
   in all four cases and substantially different from estimates based on CO 
   observations.
}
\maketitle

\newcommand{\kms}{km~s$^{-1}$}
\newcommand{\cm}{cm$^{-2}$}
\newcommand{\htwo}{\ensuremath{{\rm H_2}}}
\newcommand{\nhtwo}{\ensuremath{N_{\rm H_2}}}
\newcommand{\noh}{\ensuremath{N_{\rm OH}}}
\newcommand{\hco}{\ensuremath{{\rm HCO^+}}}
\newcommand{\nhco}{\ensuremath{N_{\rm HCO^+}}}
\newcommand{\Dla}{ Damped Lyman-$\alpha$ }
\newcommand{\dla}{ damped Lyman-$\alpha$ }
\newcommand{\beq}{\begin{equation}}
\newcommand{\eeq}{\end{equation}}
\newcommand{\noi}{\noindent}
\newcommand{\lb}{\left(}
\newcommand{\dV}{\Delta V}
\newcommand{\rb}{\right)}

\section{Introduction}
\label{sec:intro}

Molecular hydrogen (${\rm H_2}$) is the primary constituent of the molecular
component of the interstellar medium and plays a crucial role in
determining the evolution of the ISM as well as  the star formation rate in
galaxies. For example, in the Milky Way, $M_{\rm H_2} \sim 5 \times 10^9 M_\odot$,
comparable to the mass of the atomic component. Since it is difficult 
to directly detect \htwo, its column density, \nhtwo, is usually 
inferred from observations of other species; these are referred to as {\it tracers} 
of ${\rm H_2}$  (see e.g.~\cite{combes99} for a review). The most commonly used 
tracer of ${\rm H_2}$ is ${\rm CO}$, which is the second most abundant molecule 
in the ISM. Unfortunately, despite the widespread use of CO as a tracer of \htwo, 
deducing \nhtwo~from CO observations remains a fairly tricky exercise (see e.g. 
Liszt \& Lucas 1998, for a discussion).

	The OH column density is known to correlate with the visual extinction 
$A_V$ and, hence, with the {\it total} hydrogen column density, $N_{\rm H}$ 
(\cite{crutcher79}). Lucas \& Liszt (1996) and Liszt \& Lucas (1998, 1999) 
examined the variation of OH and other species (including ${\rm H_2CO}$, 
HCN, HNC and ${\rm C_2H}$) with HCO$^+$ and found that most molecules (except OH) 
showed a non-linear dependence on $N_{\rm HCO^+}$, with a rapid increase in 
their abundances at \nhco~$\approx 10^{12}$~\cm. However, \noh~and \nhco~were 
found to have a linear relationship extending over more than two orders of 
magnitude in \nhco (\cite{liszt96}), with 
\begin{equation}
\label{eqn:hco-oh}
\frac{N_{\rm HCO+}}{N_{\rm OH}} \approx 0.03 \; \; .
\end{equation}
\noindent Further, the relative abundances of OH and HCO$^+$ to \htwo~and each 
other were found to be constant in a variety of galactic clouds (\cite{liszt99}), 
with ${N_{\rm OH}} / {N_{\rm H_2}} \approx 1 \times 10^{-7}$. Based on these 
observations, Liszt \& Lucas (1999)  suggested that OH and HCO$^+$ were 
good tracers of \htwo.

	There are presently four known molecular absorption line systems at intermediate 
redshifts ($z \sim 0.25 - 0.9$) with detected HCO$^+$ (Wiklind \& Combes 1995, 1996a, 
1996b, 1997).  Until recently, OH absorption had been detected in only one of these 
objects, the $z \sim 0.886$ absorber 
towards PKS~1830-211 (\cite{chengalur99}). We have now carried out a deep search for 
redshifted OH absorption in the remaining three 
absorbers with the GMRT, resulting in detections of absorption in all cases. In this 
letter, we describe our GMRT observations of two of these absorbers, viz.
PKS~1413+135 ($z = 0.2467$) and B3~1504+377 ($z = 0.6734$); the OH obervations
of B~0218+357 are discussed in Kanekar et al. (2001). We also compare the 
OH column densities obtained in the four absorbers with their HCO$^+$ column 
densities and find that the linear relationship between OH and HCO$^+$ found in 
the Milky Way persists out to moderate redshifts. Finally, we use the conversion 
factor suggested by Liszt \& Lucas (1999) to estimate $\nhtwo$ in all these 
absorbers. Throughout this paper, we use $H_0 = 75$~km~s$^{-1}$~Mpc$^{-1}$ and 
$q_0 = 0.5$.  

\section{Observations and Data analysis}
\label{sec:obs}

	The GMRT observations of PKS~1413+135 and B3~1504+377 were carried out in 
June and October 2001, using the standard 30-station FX correlator. This provides 
a fixed number of 128 spectral channels over a bandwidth which can be varied between
64 kHz and 16 MHz. We used a 4~MHz bandwidth for B3~1504+377, thus including both 
the 1665 and 1667 MHz OH transitions in the same band and yielding a resolution 
is $\sim 9.4$~\kms. However, in the case of PKS~1413+135, the HCO$^+$ and other 
millimetre lines have very narrow widths. We hence used a bandwidth of 1 MHz and only 
observed 
the 1667 MHz transition (the stronger of the two lines, in thermal equilibrium),
with a resolution of $\sim 1.75$~\kms. The standard amplitude calibrators 3C48, 
3C286 and 3C295 were used for both absolute flux and system bandpass calibration 
in both cases. No phase calibration was necessary as both PKS~1413+135 and 
B3~1504+377 are unresolved on even the longest baselines of the GMRT. Only 
thirteen and seventeen antennas were used for the final spectra of B3~1504+377 
and PKS~1413+135, respectively, due to various maintenance activities and debugging; 
the total on-source times were 6 hours and 5.5 hours respectively.

        The data were analysed in AIPS using standard procedures. Continuum 
emission was subtracted using the AIPS task UVLIN; spectra were then extracted 
in both cases by simply averaging the source visibilities together, using the AIPS 
task POSSM, since, as mentioned above, both sources are phase calibrators for the 
GMRT. Finally, the fluxes of B3~1504+377 and PKS~1413+135 were measured to be 1.2~Jy 
and 1.6~Jy respectively; our experience with the GMRT indicates that the flux 
calibration is reliable to $\sim 15$\%, in this observing mode.

\begin{figure*}
\epsfig{file=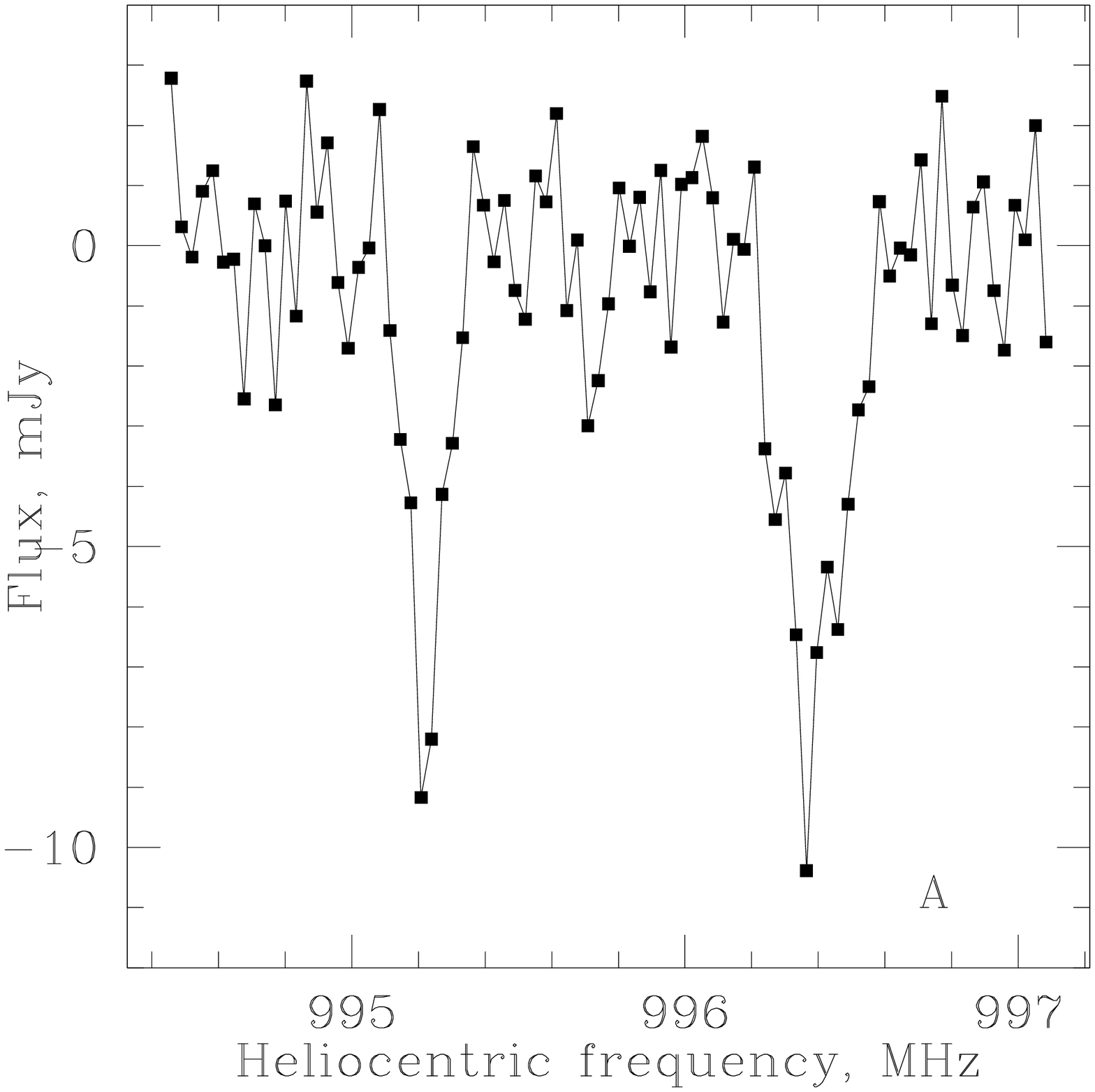,width=2.4in, height=2.4in}
\vskip -2.4 in
\hskip 2.4 in \epsfig{file=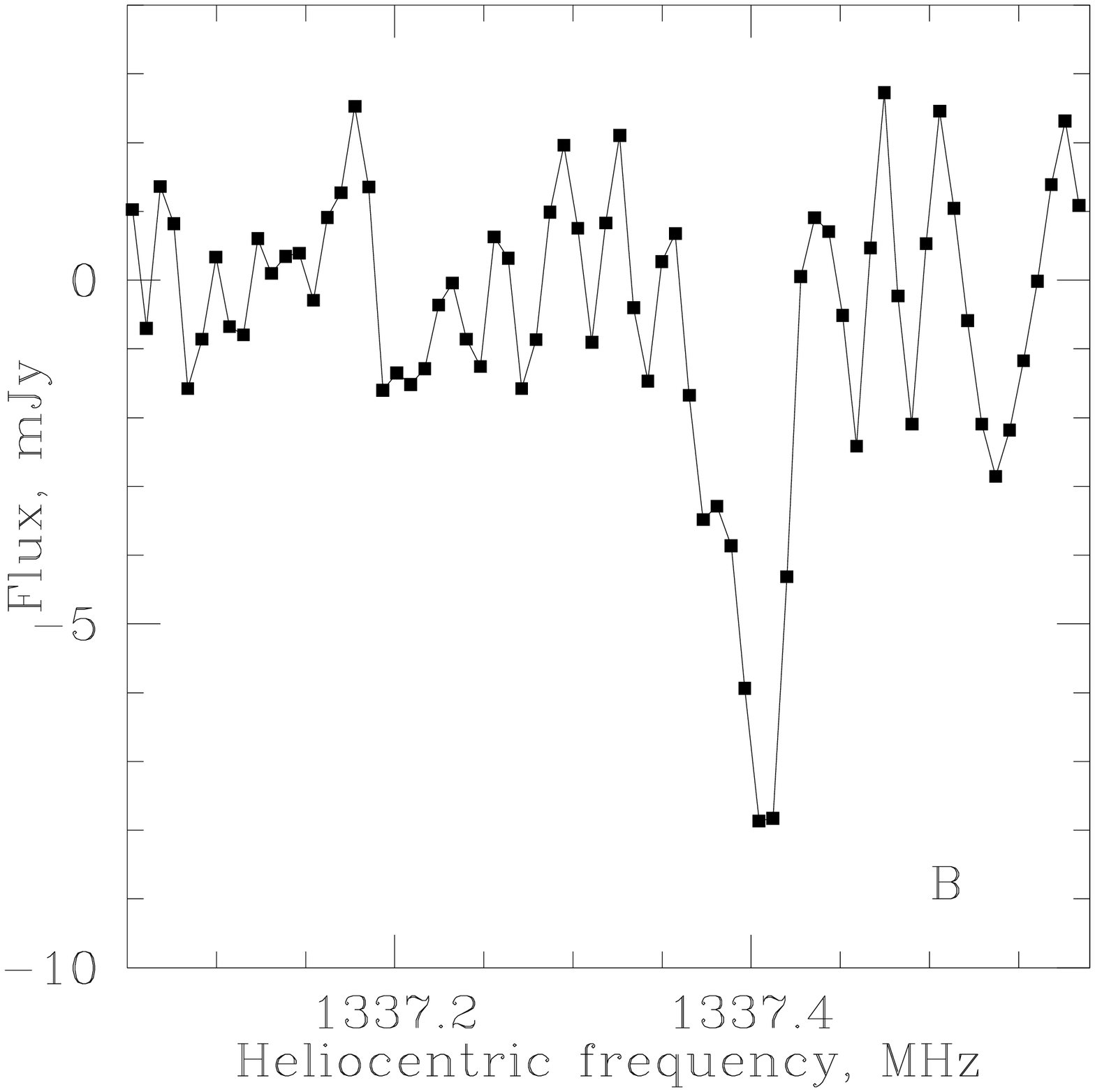,width=2.4in,height=2.4in}
\vskip -2.4 in
\hskip 4.8 in \epsfig{file=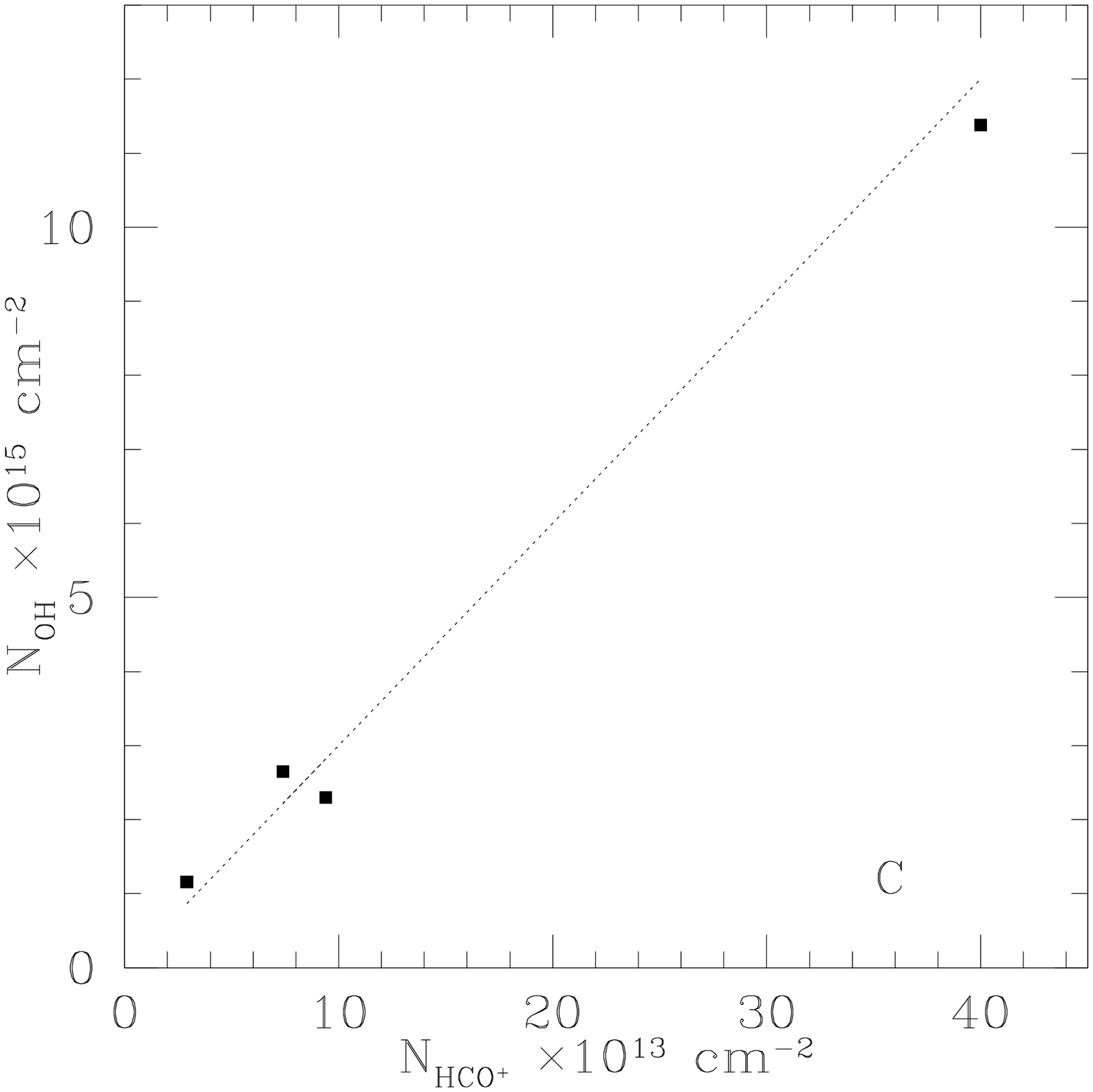,width=2.4in,height=2.4in}
\caption{ [A] 9.4~\kms resolution OH spectrum towards B3~1504+377. The spectrum 
              includes the 1665 \& 1667 OH lines.
          [B] 3.5~\kms resolution spectrum of 1667 OH line towards PKS~1413+135.
          [C] $\noh$ versus $\nhco$ for the four absorbers of our sample. The 
	      dotted line shows the correlation found in molecular clouds in the 
	      Milky Way, and is not a fit. Note that the uncertainty in \noh~is 
	      dominated by the uncertainties in $T_x$ and the covering factor $f$; 
	      the statistical errors on the optical depth are small ($< 10\%$) in 
	      all cases}
\label{fig}
\end{figure*}

The final GMRT 4 MHz spectrum towards B3~1504+377 is shown in Fig.~\ref{fig}[A].
No smoothing has been applied; the RMS noise is 1.3~mJy per 9.4~\kms~channel. Two 
absorption lines can be clearly seen in the spectrum, centred at heliocentric 
frequencies of 995.208 MHz and 996.365 MHz. These correspond to the 1665.403~MHz 
and 1667.359~MHz transitions of OH, with redshifts $z_{1665} = 0.67342 
\pm 0.00003$ and $z_{1667} = 0.67344 \pm 0.00003$. Note that the redshifts of the 
lines agree, within our error bars; we will use $z = 0.67343$, the average of 
the redshifts of the two lines, as the redshift of the OH absorption (but see
also the discussion below). The peak line depths are 9.1 mJy and 10.0 mJy, 
implying peak optical depths of $\sim 0.76$~\% and $0.83$~\% for the 1665 
and 1667 MHz transitions respectively.

The final GMRT 1 MHz spectrum towards PKS~1413+135 is shown in Fig.~\ref{fig}[B]. 
The spectrum has been Hanning smoothed and has an RMS noise of 1.3~mJy per 
3.5~\kms~channel. Absorption can again be clearly seen, with a peak line flux of 
7.9~mJy at a heliocentric frequency of 1337.404~MHz. This can be identified with the 
1667.359 MHz OH line, with a redshift $z = 0.24671 \pm 0.00001$. The peak 
optical depth is $0.49$~\%.

\section{Discussion}
\label{sec:dis}

\begin{table*}
\begin{center}
\caption{Summary of OH absorption studies}
\label{table:noh}
\begin{tabular}{|c|c|c|c|c|c|c|c|}
\hline
Source& $z_{abs}$ & \noh & \nhco$^{\star}$& \nhco$^{\dagger}$& \nhtwo$^{\ddagger}$ & \nhtwo$^{\ast}$ & $A_V$\\
&&&  & {\it (Obs.)} & (OH) & (CO) & \\
& & $10^{15}$~\cm & $10^{13}$~\cm & $10^{13}$~\cm & $10^{22}$~\cm & $10^{22}$~\cm & \\
&&&&&&& \\
\hline
&&&&&&& \\
PKS~1413+135 & 0.24671 & 1.16  & 3.5  & 2.9  & 1.16  & 0.04 & 17.8  \\
B3~1504+377  & 0.67343 & 2.3   & 6.9  & 9.4  & 2.3   & 0.12 & 27.1  \\
B~0218+357   & 0.68468 & 2.65  & 7.8  & 7.4  & 2.65  & 40.0 & 28.9  \\
PKS~1830-211 & 0.88582 & 11.38 & 34.2 &  40  & 11.38 & 4.0  & 123.2 \\
&&&&&&& \\
\hline
\end{tabular}
\end{center}
\vskip 0.05in
${}^\star$~Obtained using equation~\ref{eqn:hco-oh}.\\
${}^\dagger$~Actual measured HCO$^+$ column densities, from Wiklind \& Combes 1995, 
1996a, 1997, 1998 and Menten et al. 1999. \\
${}^\ddagger$~Obtained using $\nhtwo = 1.0 \times 10^7 \times \noh$. \\
${}^\ast$~Estimated from CO observations, from Wiklind \& Combes 1996a, 
1996b, 1997, 1999. \\
\vskip 0.01in
\end{table*}

For an optically thin cloud in thermal equilibrium, the OH column density of
the absorbing gas \noh~is related to the excitation temperature $T_x $ and the
1667 MHz optical depth $\tau_{1667}$ by the expression (e.g. \cite{liszt96})
\begin{equation}
N_{\rm OH} = 2.24 \times 10^{14} {\lb {\frac {T_x }{f}} \rb}\int \tau_{1667} 
 \mathrm{d} V \; ,
\label{eqn:noh}
\end{equation}
\noi where $f$ is the covering factor of the absorber. In the above, \noh~ is
in cm$^{-2}$, $T_x $ in K and $\mathrm{d}V$~in \kms. VLBI observations, when
available, can be used to constrain the extent of the background radio
continuum, and hence to estimate the covering factor $f$. Unfortunately,
the OH excitation temperature $T_x $ cannot be directly estimated, for
cosmologically distant objects. In the Galaxy, OH emission studies have shown
that this temperature may be as low as $T_x \sim T_{CMB} + 1$~K, with similar
values for the HCO$^+$ line ($T_x \, (HCO^+) \sim T_{CMB}$; \cite{lucas96}).
However, the excitation temperatures of the redshifted HCO$^+$ lines in three
of the four absorbers have been found to be {\it higher} than $T_{CMB}(1 + z)$, the
redshifted CMB temperature; it is thus quite likely that the OH excitation
temperature too will be higher in these systems. Given that all four absorption 
systems are believed to originate either in spiral disks or in late-type galaxies 
(\cite{lehar2000}; \cite{stocke92}; \cite{stickel94}), we will, in the absence 
of additional information, assume $T_x = 10$~K, a typical temperature in dark 
clouds in the Milky Way.

\noindent {B3~1504+377 :} The OH absorption redshift, $ z = 0.67343$, is 
in good agreement with that of the HI ($z = 0.67340$; \cite{carilli97}); 
however, the molecular absorption seen in the millimetre wave bands peaks about
15~\kms~away, at $z = 0.67335$ (\cite{wiklind96a}). Next, while the two 
$z = 0.67343$ OH lines are not very different in peak optical depth, the 
1667~MHz line can be seen to be the wider of the two (total spread $\sim 
103$~\kms~against $\sim 75$~\kms); these widths are similar to those of 
the $z = 0.67335$ mm-wave lines (total spread $\sim 100$~\kms;~\cite{wiklind96a}). 
The integrated optical depths of the OH lines are $\int \tau_{1665} \mathrm{d}V 
= 0.257$~\kms~and $\int \tau_{1667} \mathrm{d}V 
= 0.448$~\kms. We note that millimetre wave molecular absorption has also been 
detected at $z = 0.67150$ (\cite{wiklind96a}); the 1665 MHz OH line corresponding 
to this redshift arises at a heliocentric frequency of 996.352~MHz, and 
may thus overlap with the 1667~MHz line of the $z = 0.67343$ absorber. We will 
hence use the integrated optical depth in the 1665~MHz line of the $z = 0.67343$ 
absorber to evaluate its OH column density (assuming thermal equilibrium, i.e.
$\int \tau_{1667} \mathrm{d}V$ / $\int \tau_{1665} \mathrm{d}V = 1.8$). 
The integrated 1665 MHz optical depth then yields an OH column density 
$\noh = 1.04 \times ({T_x }/{f}) \times 10^{14}$~\cm. 
Carilli et al. (1997) estimate $f \ge 0.46$, from VLBI observations at 1.6 and 
5~GHz, with the lower value obtained if only the compact core of the radio 
continuum (size $\approx 7.2$~pc) is covered by the absorbing cloud. On the 
other hand, $f = 0.74$, if the radio jet of the source is also covered; this 
would require a cloud of size greater than $\sim 54$~pc. Typical sizes of 
Giant Molecular Clouds in the Milky Way range from 10 to 50~pc (\cite{blitz90}); 
we will hence use a covering factor $f = 0.46$ in the analysis. This yields 
$\noh = 2.3 \times 10^{15} ({T_x }/{10}) ({0.46}/{f})$~\cm.

\noindent {PKS~1413+135 :} The redshift of the OH absorption towards 
PKS~1413+135 is in excellent agreement with that of the millimetric absorption 
($z = 0.24671$; \cite{wiklind97}). The width of the 1667 MHz line is also quite 
narrow, with a total spread $\sim 14$~\kms~(slightly wider than the mm lines 
of Wiklind \& Combes (1997), which have a total spread $\la 10$~\kms); the 
integrated optical depth is $\tau_{1667} = 
0.023 \pm 0.001$~\kms. Equation~\ref{eqn:noh} then yields an OH 
column density $\noh = 0.51 \times ({T_x }/{f}) \times 10^{14}$~\cm. VLBA 
maps of PKS~1413+135 at 3.6, 6, 13 and 18~cm (\cite{perlman96}) have shown 
that the core (component N in their maps) is highly inverted, with a spectral 
index $\alpha = + 1.7$. Extrapolating their flux measurements yields a 
core flux of $\sim 70$~mJy at the GMRT observing frequency, within $\sim 3$~mas 
(i.e. $\sim 10$~pc at $z = 0.247$). Since this is likely to be covered by 
the molecular cloud, we obtain a {\it lower} limit on the covering factor, 
$f \geq 0.044$. On the other hand, if components C and D ({\cite{perlman96}) 
are also covered, it would imply $f \sim 0.2$ (using extrapolated fluxes of 
components C and D). This is, however, unlikely, given that components C, D 
and the core are spread over $\sim 20$~mas (i.e. $\sim 65$~pc at $z = 0.247$), 
larger than the size of a typical molecular cloud. We will use $f = 0.044$ 
in the analysis (note that $f \sim 0.1$ is also possible); this yields 
$\noh = 1.16 \times 10^{15} ({T_x }/{10}) ({0.044}/{f})$~\cm.

We also evaluate $\noh$ for the other two high redshift molecular absorbers
in which OH absorption has been detected, towards B~0218+357 and PKS~1830-211
(\cite{chengalur99,kanekar01}). In the case of PKS~1830-211, the integrated 
optical depth in the 1667 MHz line is $\int \tau_{1667} \mathrm{d} V = 1.83$~\kms; 
i.e. $\noh = 4.1 \times 10^{14} ({T_x /f})$~\cm. The millimetric absorption 
is known to occur only towards the south-west component of the background 
source, which contains $\sim 36$~\% of the radio flux (\cite{wiklind98}); 
the covering factor is thus likely to be $f \sim 0.36$. We then obtain 
(again using $T_x = 10$~K) $\noh = 11.4 \times 10^{15} ({T_x / 10}) 
({0.36 / f})$~\cm. Similarly, the integrated optical depth in the 1667 MHz line 
is $\int \tau_{1667} \mathrm{d}V = 0.772$~\kms, in the case of the $z = 0.6846$ 
absorber towards B~0218+357 (\cite{kanekar01}); thus, $\noh = 1.1 \times 10^{14} 
\times ({T_x / f})$~\cm. Carilli et al. (1993) estimate the covering 
factor to be $f \sim 0.4$, assuming that only component A of the background 
continuum source is covered by the absorbing cloud. The latter is reasonable 
since it is known that the millimetric absorption also only occurs against this
component (\cite{wiklind95}). The OH column density is then $\noh = 2.65 
\times 10^{15} ({T_x / 10}) ({0.4 / f})$~\cm.

Figure~\ref{fig}[C] shows a plot of the OH column density versus the HCO$^+$ 
column density for the four absorbers of our sample. The dotted line 
is the relationship found in the Milky Way. All four systems lie close
to this line;  the linear relationship between OH and HCO$^+$ clearly appears to
persist out to 
moderate redshifts. Table~\ref{table:noh} summarises our results and 
also lists the $\htwo$ column densities (evaluated using $\noh / \nhtwo = 10^{-7}$) 
for the four absorbers of our sample; these can be seen to be quite different 
from the values estimated from CO observations (penultimate column).
It is interesting that our estimate of the $\htwo$ column density in the 
$z = 0.6846$ absorber towards B~0218+357 is in reasonable agreement with that 
obtained by Gerin et al. (1997) ($\nhtwo = 2 \times 10^{22}$~\cm), using the 
${}^{17}{\rm CO}$ line. We also note that the good agreement between the observed 
HCO$^+$ column densities and those estimated using equation~\ref{eqn:hco-oh}
is despite the fact that we have used the general excitation temperature, 
$T_x = 10$~K, for the OH line in all cases. For example, the HCO$^+$ excitation 
temperature is measured to be 13~K, in the case of the $z = 0.6734$ absorber 
towards B3~1504+377; if this value were also used for $T_x $, one would obtain 
$\nhco = 9.0 \times 10^{13}$~\cm, in even better agreement with that obtained 
from the HCO$^+$ absorption spectra of Wiklind \& Combes (1996a).

Finally, the last column of table~\ref{table:noh} gives the visual extinction 
along the four lines of sight, evaluated using the $\htwo$ column densities of 
column~8, the HI column densities obtained assuming a spin temperature of 100~K
(\cite{carilli92}; \cite{carilli93}; \cite{carilli97}; \cite{chengalur99}),
and a Galactic extinction law ($R_V = 3.1$; \cite{binney98}). We note that 
a far lower extinction ($A_V \sim 28$) is obtained towards B~0218+357 
than that estimated from observations of the ${}^{12}{\rm CO}$ line ($A_V 
\ga 500$; \cite{wiklind99}). While $A_V = 28$ is still quite large and does 
require the presence of fine structure in the molecular cloud (since component~A 
is visible in the optical; \cite{wiklind99}), the present value requires a less 
dramatic change in physical conditions across the optical source than that obtained 
by Wiklind \& Combes (1999).  We also find a high extinction towards PKS~1413+135 
($A_V \sim 18$), although still somewhat smaller than that obtained from the 
deficit of soft X-rays ($A_V \ga 30$; \cite{stocke92}).

In summary, we find that the linear relationship between OH and HCO$^+$ column 
densities, seen in Galactic molecular clouds, appears to persist out 
to absorbers at intermediate redshift, with $\nhco \approx 0.03 \times \noh$.
One may thus be able to use OH absorption lines to trace the $\htwo$ content 
of molecular clouds at cosmological distances; all four absorbers of our sample 
have $\nhtwo \ga 10^{22}$~\cm.

\vskip 0.05 in
\noi {\bf Acknowledgments} The GMRT observations presented in this paper 
would not have been possible without the many years of dedicated effort 
put in by the GMRT staff to build  the telescope. The GMRT is run by the
National Centre for Radio Astrophysics  of the Tata Institute of Fundamental 
Research. We thank Chris Carilli for illuminating discussions, which were 
useful in planning the observations.

\end{document}